# Dynamical Networks of Influence in Small Group Discussions

**Mehdi Moussaïd[1*], Alejandro Noriega Campero[2] and Abdullah Almaatouq[2]**

[1] Center for Adaptive Rationality, Max Planck Institute for Human Development, Berlin, Germany.

[2] Massachusetts Institute of Technology, Cambridge, MA, USA.

*Corresponding author: moussaid@mpib-berlin.mpg.de

# Abstract

In many domains of life, business and management, numerous problems are addressed by small groups of individuals engaged in face-to-face discussions. While research in social psychology has a long history of studying the determinants of small group performances, the internal dynamics that govern a group discussion is not yet well understood. Here, we rely on computational methods based on network analyses and opinion dynamics to described how individuals influence each other during a group discussion. We consider the situation in which a small group of three individuals engages in a discussion to solve an estimation task. We propose a model describing how group members gradually influence each other and revise their judgments over the course of the discussion. The main component of the model is an influence network — a weighted, directed graph that determines the extent to which individuals influence each other during the discussion. In simulations, we first study the optimal structure of the influence network that yields the best group performances. Then, we implement a social learning process by which individuals adapt to the past performance of their peers, thereby affecting the structure of the influence network in the long run. We explore the mechanisms underlying the emergence of efficient or maladaptive networks and show that the influence network can converge towards the optimal one, but only when individuals exhibit a social discounting bias by downgrading the relative performances of their peers. Finally, we find a late-speaker effect, whereby individuals who speak later in the discussion are perceived more positively in the long run and are thus more influential. The numerous predictions of the model can serve as a basis for future experiments, and this work opens research on small group discussion to computational social sciences.

# Introduction

In many domains of life, complex problems can be successfully addressed by pooling the knowledge of several individuals [1,2]. When making decisions, forming judgments, or solving multidimensional problems, groups of people can outperform the best individual in the group, and sometimes even the experts in the problem domain. In everyday life, this collective achievement is commonly accomplished by means of face-to-face group discussions, during which the exchange of information and ideas between people results in the emergence of accurate collective solutions [3]. Whereas research in social psychology has a long history in studying the performances of small group discussions, more recent methods of computational social science are less often used to address this issue [4–7]. In this context, the present article introduces a network approach to study the internal dynamics that operate during a group discussion.

Given the omnipresence of group discussions in many areas of life, the factors impacting the performances of a group discussion have been extensively studied in the past. Classical research on group performance has highlighted numerous detrimental effects that can impair the quality of the discussion [3]. For instance, the hidden profile effect refers to the situation where group members fail to share important private information and tend to focus mostly on the elements of information known by the majority of them [8,9]. Likewise, groupthink and conformity are common issues that arise during discussions and occur when the group members ignore important facts or unwillingly adopt the judgment of others to reach a non-contentious collective consensus [10,11]. Also, group discussions can be subject to polarization effects, in which the judgments of the individuals tend to become more extreme as a result of social interactions [12,13]. Nevertheless, group discussions remain a powerful mean to aggregate the ideas and judgments of several people. In controlled experimental settings, it has been shown many times that groups can outperform single individuals in a wide variety of tasks, such as for detecting lies [14], reconstructing noisy signals [15], establishing a medical diagnosis [16], and in a variety of binary-choice tasks [17].

Yet, the conditions under which a group would perform good or bad remain unclear. In a recent series of experimental studies, Woolley et al. revealed the existence of a 'collective intelligence factor' that is predictive of groups performance across a wide variety of tasks [1]. That factor is not associated with the average skills of the individual group members. Rather, it strongly correlates with the social sensitivity of the individuals, that is, their ability to listen and integrate the arguments of the others, and to balance the speaking turns across all group members. This suggests that one key aspect of group performance lies in the *internal dynamics* that operate during the discussion, more than in the individual skills of the group members. However, although the collective intelligence factor is a powerful indicator to anticipate the group's performance, it does not explain the underlying causal mechanisms leading to collective good or bad performances. In fact, the *dynamics* of the group discussion, that is, the pattern of communication that takes place during the discussion and the social influences that operate among group members is not yet well understood.

This dynamical aspect of collective intelligence has been deeply investigated in a different domain. In the past decade, computational social scientists have begun to understand more precisely the dynamics driving judgment formation and social contagion in large populations of people composed of hundreds of individuals connected in social networks [18–20]. Numerical models have been proposed to describe how repeated interactions between a large number of individuals can possibly drive a population towards a consensual judgment, or on the contrary, polarize the beliefs of the crowd [13,21–23]. These models generally rely on the assumption that agents tend to revise their judgments by averaging their own and their neighbors' judgments, gradually converging towards a consensus. A similar averaging process has also been used in numerous models of advice-taking in psychology, this time at the scale of a dyad [24,25]. Nevertheless, most existing research of opinion dynamics has dealt with large social networks, often focusing on how the network topology impacts the propagation of judgments. However, these methods have rarely been applied to the case of face-to-face discussions, where the group size is small — typically three to five individuals — and where all the individuals are interconnected in a full network.

In the present work, we aim at describing the internal dynamics that operate during group discussions, using tools and concepts inherited from the network science and the computational social sciences. For this, we describe the group as a small social network in which each group member is represented by a node, and all the nodes are connected to one another by weighted ties representing the extent to which individuals influence each other. In simulations, we show that the structure of this influence network determines the performance of the group during a group discussion. Importantly, we also assume that individuals can adapt the weight they assign to their peers after observing their past performances: Good performers tend to become more influential, and bad performers tend to lose influence in the group. Over time, the influence network evolves and often converge to the optimal structure. Crucially, this only happens when individuals exhibit a social discounting bias, that is, when people systematically downgrade the relative performances of their peers. Finally, we show that the speaking order has significant consequences on the emerging structure of the influence network, thus drawing links to the collective intelligence factor. The surprisingly complex dynamics that emerge from our simple model opens numerous experimental perspectives for future research.

## Model

**Discussion dynamics**. Our model describes the process of group discussions, in which *N* individuals undertake an estimation task collectively. Each individual *i* in the group has an initial estimate $x_i^0$ drawn from a normal distribution with mean $\mu$ and standard deviation $\sigma$. The discussion is composed of $N_r$ speaking rounds across which the individuals progressively revise their initial estimate. The estimate of the individual *i* at round *r* is noted $x_i^r$. In each speaking round *r*, a randomly selected individual speaks up and communicates her current estimate $x_i^r$ to all the others. Every time an individual speaks up, all the other group members revise their current estimate using a weighted average procedure (see, e.g., [13,24,25]. Formally, the revised estimate of the individual *j* after the individual *i* has spoken up is given by

$$x_j^r = x_j^{r-1} + w_{ij} \, (x_i^r - x_j^{r-1}) \, .$$

In the above equation, the term $w_{ij}$ represents the weight that the individual *j* assigns to the speaker *i*. The weight is defined in the interval $[0 \ 1]$. According to the above equation, a weight $w_{ij} = 0$ indicates that *j* ignores the judgment of *i*, and a weight $w_{ij} = 1$ indicates that *j* fully adopts the judgment of *i*. The speaker does not revise her estimate in round $r$, leading to $x_i^r = x_i^{r-1}$. The same process repeats round after round, until the last round $r = N_r$.

The weights $w_{ij}$ are not necessarily the same for all pairs of individuals and the weight $w_{ij}$ is not necessarily identical to $w_{ji}$. Hence, the $N$ individuals are connected by an *influence network*, that is, a weighted directed graph that determines how group members influence one another during the discussion. The **figure 1** illustrates the dynamics of a group discussion for two different influence networks.

The above equation of social influence has been experimentally confirmed and used in numerous models of opinion dynamics (see, e.g. [12,13,21,23,25,26]). Note that, in principle, the weight factors $w_{ij}$ do not need to be bounded to the interval $[0 \ 1]$. Weights higher than 1 or lower than 0 could represent more extreme social influence phenomena, such as social repulsion ($w_{ij} < 0$) or over-adoption ($w_{ij} > 1$) — which have potential to generate group polarization [27]. Nevertheless, we choose to restrict ourselves to weights varying in the interval $[0\ 1]$ in the present study for simplifying the traceability of the simulation results. Another simplification of our model is that, in contrast to other formalizations [28], the weights are associated with a given *person* and not to a given *argument* that a person formulates. The model, therefore, assumes that some individuals are naturally more influential than others, rather than considering the persuasiveness of each communicated argument separately.

Our approach differs from the simple averaging of the initial estimates that are typically used in the "wisdom-of-crowds," and from the repeated averaging across all group members typically used in a DeGroot updating procedure [29]. Here, individuals only integrate the estimate of the last speaker and do not average across all individuals simultaneously. This creates complex dynamics involving judgment propagation and indirect influence among group members. In the first part of the 'Results' section, we study the optimal structure of the influence network for various group compositions.

**Social learning**. Our model does not only focus on the outcome of the discussion but also on how individuals adapt to it in the long run. For this, we assume that the *same* group of individuals undertakes not only one, but a series of $N_T$ estimation tasks from the same problem domain. For each estimation task, a new discussion takes place between the same set of individuals, following the procedure described in the previous section. For the first discussion, the group members are strangers and know nothing about each other's skills. However, as individuals undertake repeated estimation tasks together, they can learn about and adapt to each other's past performances. This social learning aspect is represented by a change in the weights that each individual gives to the others. In other words, the influence network evolves over time, depending on how the individuals perceive their peers.

Formally, we now include a time dependency on the weights $w_{ij}(t)$, where the variable $t$ varies from 1 to $N_T$. The variable $t$ indicates the number of discussions that the pair of individuals $\{i,j\}$ undertook together. Individuals who had no past interactions with their partner assign a default weight $w_{ij}(0) = w^0$ to him or her.

Previous experimental measurements have shown that individuals update the weight assigned to others based on their relative, not absolute, performances [30]. Furthermore, experimental data have also revealed the existence of a social discounting bias in this process, indicating that people tend to underweight their own error as compared to the errors of their partners [25,30,31]. In our model, we describe these facts by assuming that the weight given by *i* to *j* is increased by an offset $w^*$ if *j* performed sufficiently better than *i* during the previous discussion, and is decreased by $w^*$ otherwise:

$$w_{ij}(t) = w_{ij}(t-1) + w^* \quad \text{if } e_j^* + \alpha < e_i^0$$

And

$$w_{ij}(t) = w_{ij}(t-1) - w^* \quad \text{if } e_j^* + \alpha > e_i^0$$

Here, $e_i^0 = |x_i^0|$ is the error that the focal individual *i* made on her initial estimate $x_i^0$ during the previous discussion, and $e_j^* = |x_j^*|$ is the error that the individual *j* committed on the first communicated estimate during the previous discussion. This formalization reflects the fact that the focal individual *i* does not know what was the initial estimate $x_j^0$ of the individual *j*, and can only consider the first *communicated* estimate $x_j^*$ of the individual *j* to judge him or her. The parameter $\alpha$ is the social discounting bias. The higher $\alpha$ the stronger *i* downgrades the quality of *j*'s judgments.

In the second part of the 'Results' section, we explore how the weights $w_{ij}(t)$ — and thus the structure of the influence network—evolve as *t* increases, and compare the emerging group structure to the optimal one.

# Results

**Optimal group configuration**. Ignoring the social learning aspect of the model for now (i.e. considering $N_T = 1$), we addressed the question of what are the optimal weights $w_{ij}$ that each individual should assign to all the others such that the group error is minimized. Is the group better off by assigning equal weights to everybody, irrespective of the individual members' skills, or is it more efficient to give a stronger power to the best performers?

To address this question, we varied the group composition by defining two types of individuals: 1) the good performers, for whom the initial estimates are drawn from a normal distribution with mean $\mu$ and standard deviation $\sigma^+$; and 2) the bad performers, for whom the initial estimates are drawn from a normal distribution with the same mean $\mu$ but a standard deviation $\sigma^- > \sigma^+$ (**Figure S6**). We defined the group error $E$ as the average error of the group members at the end of the discussion: $E = \sum_i e_i^{N_r}/N$, where $e_i^{N_r} = |x_i^{N_r}|$ is the final error of the individual *i*.

Using an optimization procedure (see the Methods section), we computed the optimal network structure — that is, the weight values $w_{ij}$ for all pairs $\{i,j\}$ — that minimizes the final error $E$ of the group for different group compositions. The results are presented in **Figure 2**. Groups composed of equally skilled members (either good or bad performers) reach their best performances when individuals assign an equal weight $w \approx 0.2$ to each other. When individuals do not perform equally, however, the weights need to be adjusted accordingly. For instance, in groups composed of two good and one bad performer, the group performs best when the two good performers assign a weight $w_{ij} \approx 0.2$ to one another while ignoring the bad performer, but at the same time receiving a weight $w_{ij} \approx 0.7$ from her. In the next section, we study whether groups can naturally converge towards these optimal structures via social learning.

**Emerging patterns**. Next, we addressed the question of whether groups can self-organize to reach the optimal structures described in **Figure 2**. For this, we conducted another set of simulations, this time allowing for social learning across a series of $N_T = 100$ discussions. For each group composition, we also varied the value of the social discounting bias $\alpha$ in the interval $[0\ 2]$. For all values of $\alpha$, we measured the average group performance after $N_T = 100$ discussions, for different group compositions. Surprisingly, we found that the best collective performances are found for a social discounting bias $\alpha > 0$ (**Figure 3**). That is, individuals do benefit from moderately downgrading the performances of their peers. To better understand this result, we looked at the associated network structures for three values of $\alpha$ ($\alpha = 0$, $\alpha = 0.1$, and $\alpha = 1$). The results are shown in **figure 4**. It is visible from this figure that in the absence of bias (i.e., $\alpha = 0$) the weights that individuals assign to each other are too high. However, increasing the bias tends to reduce the overall weight values. When the social discounting bias is large enough, the weights of the influence networks match the optimal ones presented in **figure 2**, and yield the best group performances shown in **figure 3**.

Why do people benefit from downgrading the performances of their peers? Social discounting is necessary to counterbalance the fact that individuals tend to *overestimate* the skills of their peers. The reason is that individuals judge the performance of the others based on the first estimate $x_i^*$ they communicated, which is generally better than their real initial estimate $x_i^0$. For instance, in the illustrative discussion dynamics sketched in Figure 1A, the individual $p_3$ (in yellow) communicates her first estimate $x_3^*$ at round 4. This estimate is very close to the true value, giving the impression that $p_3$ is an excellent performer. However, the actual initial estimate $x_3^0$ of $p_3$ was far off. Because the initial estimate $x_3^0$ was not communicated, the other group members could only judge the performance of $p_3$ based on $x_3^*$ and thus overestimated her skills. Generally speaking, the first communicated estimate $x_i^*$ tends to be more accurate than the initial estimate $x_i^0$, because $x_i^*$ has been revised in light of what the others have communicated before [2,32]. For that reason, the weights are usually too high when $\alpha = 0$ (Figure 4). Social discounting can correct this overestimation and is therefore beneficial to the group members.

**Speaking order effect.**

One important side effect of the above mechanism is that individuals who tend to speak for the first time later in the discussion are more likely to be positively perceived by their peers. In fact, one can remain silent during the beginning of the discussion, integrate the estimates communicated by the others, and speak up later to communicate a revised and more accurate estimate to the rest of the group. This would have the effect of giving others the impression that the late speaker is a good performer. We evaluated the late-speaker effect in an additional series of simulations, by manipulating the round at which one group member speaks up for the first time. As predicted, the average weight $w_{ij}(N_T)$ that the individual $j$ receives from the others after $N_T$ discussions is significantly increased as $j$ speaks for the first time later in the discussion (**Figure 5A**). The late-speaker effect is attenuated for the calibrated values of $\alpha = 0.1$, but does not disappear completely.

This result contrasts with the empirical fact that the individual who speaks first in a group deliberation have a stronger impact on the outcome of the discussion (generally known as the anchoring bias; see e.g., [33]). Interestingly, this "first-speaker" effect is also visible from our simulations (**Figure 5B**). In fact, the first-speaker and late-speaker effects are not incompatible: On the one hand, individuals who speak early during a discussion have a stronger influence *on* that discussion. On the other hand, however, individuals who speak late during a discussion have the stronger influence in the long run, because they tend to receive greater weights from others.

# Discussion

Based on methods inspired by network science and opinion dynamics, we studied how the internal structure of a group could emerge and shape the group's collective performances. For this, we introduced the influence network — a weighted, directed graph that determines the extent to which each group member influences the others. We showed that the structure and the evolution of that influence network could be a major determinant of group performance: Groups perform well when their internal structure reflects the skills of the group members well, but perform poorly otherwise. It is also interesting to compare the performances of face-to-face discussions with those of other methods of collective intelligence such as the wisdom-of-the-crowds approach (WOC). In contrast to face-to-face discussions, the WOC computes the average estimate of the group members in the absence of any social interactions [34]. Our additional simulations (see supplementary **figure S7**) show that the WOC outperforms the group discussion when the skills of the group members are similar. However, for groups composed of a mixture of good and bad performers, the discussion outperforms the WOC on the long-run because the group members will eventually find out who are the best performers and follow them while ignoring the judgments of the bad performers. In other words, groups can adapt to the skills of the group members whereas the WOC averages across everybody's estimates irrespective of their individual skills.

In the context of group discussions, previous experimental studies have revealed the existence of a 'collective intelligence factor' — called $c$ — that is predictive of groups performance [1]. The authors of that study have shown that $c$ correlates with the social sensitivity of the individuals, as measured by the 'Reading the mind in the eyes' test [35]. That is, groups composed of individuals with higher social sensitivity tend to perform better than those with lower social sensitivity. An important question would then be whether this correlation between the social sensitivity of the individuals and the group's performance could be explained by the structure of the group's influence network. It is conceivable that individuals with a higher social sensitivity have a better ability to perceive the skills of their peers and to adjust the weights they give them during a discussion and in the long run. On the contrary, individuals with a lower social sensitivity would fail to adequately balance the weight they give to one another and produce maladaptive influence networks leading to poor collective performances.

Another important component of the collective intelligence factor is the ability of the group members to take conversational turns equally. Experiments have shown that groups where a few people dominate the conversation are outperformed by those with an equal distribution of speaking turns. In our simulations, however, all individuals have equal probability to speak up at each discussion round, and the impact of unbalanced speaking turns was not explored. The reason is that the relationship between an individual's skills, social influence, and speaking frequency is unclear. The speaking probability can be affected by the individual's skills, or by the individual's status in the influence network. This aspect of the discussion dynamics needs to be evaluated experimentally.

In sum, our simple model produces a rich set of predictions that could constitute important explanations to existing research on group discussion. This work calls for a series of experimental studies that would (1) validate the predictions, (2) test the relationship with the individual's social sensitivity, and (3) evaluate the determinants of the speaking frequency. With that regards, novel technological tools such as the sociometric badges recording people's speaking frequency could help measuring the internal dynamics that take place during a group discussion [36,37]. The model can open numerous perspectives aimed at enhancing the collective performances of groups in all situations where people engage in face-to-face discussions for solving problems or making decisions. This applies to many domains of life including the business and industry, scientific research, politics and medical decision-making.

## Methods

The optimal influence networks presented in figure 2 were computed through an exhaustive search optimization procedure. For each of the four group compositions presented in figure 2, we systematically varied the six weight values of the network in the interval [0 1], with steps of 0.2. In such a way, we tested a total of 46656 different configurations for each group composition. For each configuration, we measured the average group error across 5000 discussions. The best 30 configurations that produced the smaller group errors were then merged by averaging the weights $w_{ij}$ across them. The six resulting weights are those presented in figure 2.

## Data Availability
The simulation code of the results presented in this work is available in the supplementary material.

## Competing Interests
We have no competing interests

## Authors' Contributions
MM conceived of the study, conducted simulations, carried out the statistical analyses, and drafted the manuscript; AN participated in the design of the study and helped draft the manuscript; AA participated in the design of the study and helped draft the manuscript; All authors gave final approval for publication.

## Funding
This research was supported by a grant from the German Research Foundation (DFG) as part of the priority program on *New Frameworks of Rationality* (SPP 1516) awarded to Ralph Hertwig and Thorsten Pachur (HE 2768/7-2). The funders had no role in study design, data collection and analysis, decision to publish, or preparation of the manuscript.

## Research Ethics
No ethical assessment was required for this study.

## Animal Ethics
No ethical assessment was required for this study.

## Permission to carry out fieldwork
No permissions were required for this study.

## Acknowledgements
We thank Solal Moussaïd for fruitful discussions.

Figures

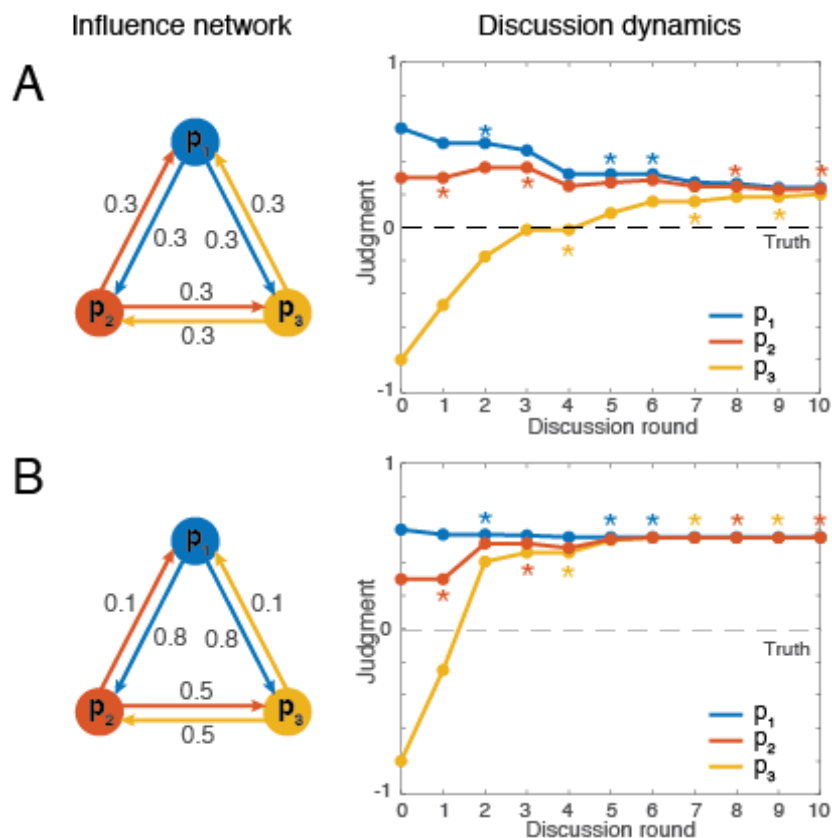

**Figure 1**: **Illustrative examples of group discussions with *N*=3 individuals**. (A) Simulation with a balanced influence network (represented on the left side), in which all group members assign the same weight $w_{ij} = 0.3$ to all other group members. In our representation of the influence network, the arrows pointing from an individual *i* to another individual *j* represent the influence that *i* has on the judgment of *j*. The corresponding discussion dynamics is depicted on the right side. The color stars indicate the identity of the speaker at each round of the discussion. The initial judgments of the three individuals for this simulation are $x_1^0 = 0.6$, $x_2^0 = 0.3$, and $x_3^0 = -0.8$. At round $r = 1$, the individual 2 (in red) speaks up and communicates her judgment $x_2^0$ to the two others. Individuals 1 and 3 revise their own judgment accordingly, leading to $x_1^1 = 0.5$ and $x_3^1 = -0.47$. After 10 rounds of discussion, the judgments of the three individuals converge around $x = 0.2$. In each round, the identity of the speaker is randomly chosen among the *N* individuals. (B) The same simulation assuming a more hierarchical influence network. In this case, the individual 1 is more influential than the two others. During the discussion, the judgments of the three individuals converge around $x = 0.6$. To facilitate the comparison between (A) and (B), the initial judgments and the sequence of speakers are identical in both examples.

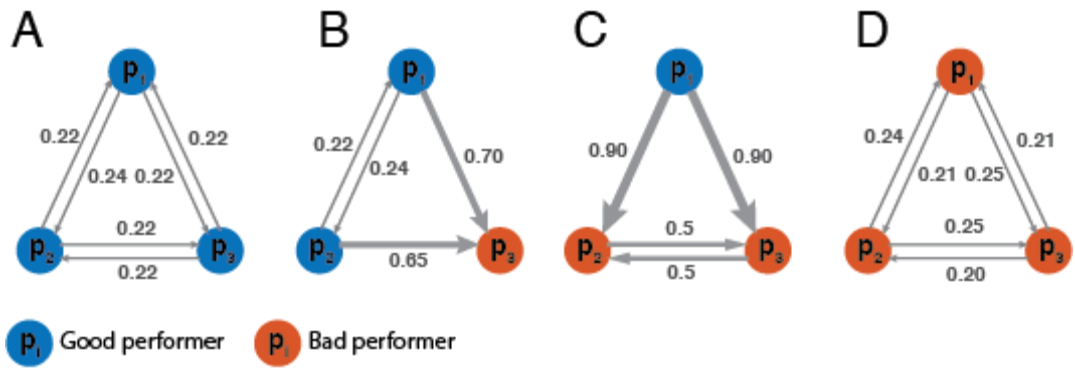

**Figure 2**: **Normative group structures**. Optimal networks of influence for different group composition and $N = 3$. Good performers draw their initial judgments from a normal distribution with mean $\mu = 0$ and standard deviation $\sigma^+ = 1$, whereas bad performers draw their initial judgments from a normal distribution with mean $\mu = 0$ and standard deviation $\sigma^- = 5$ (see **figure S6**). (A) When all group members are good performers, the best collective performance is found when all individuals assign the same weight $w_{ij} \approx 0.2$ to all others. (B) In groups composed of two good performers and one bad performer (here $p_3$ is the bad performer, depicted in red), the group performs best when the two good performers assign a weight $w_{ij} \approx 0.2$ to one another while ignoring the bad performer, but at the same time receiving a weight $w_{ij} \approx 0.7$ from her. (C) In groups composed of two bad performers and one good performer (here $p_1$ is the good performer), the best collective performance is found when $p_1$ receives a strong weight $w_{ij} = 0.9$ from the two others while at the same time ignoring them. The two bad performers give each other a weight of 0.5. (D) When all group members are bad performers, the best collective performance is found when all individuals assign the same weight $w_{ij} \approx 0.2$ to all others, similar to (A). The width of the arrows is proportional to $w_{ij}$, that is, the weight given by $j$ to the judgment of $i$.

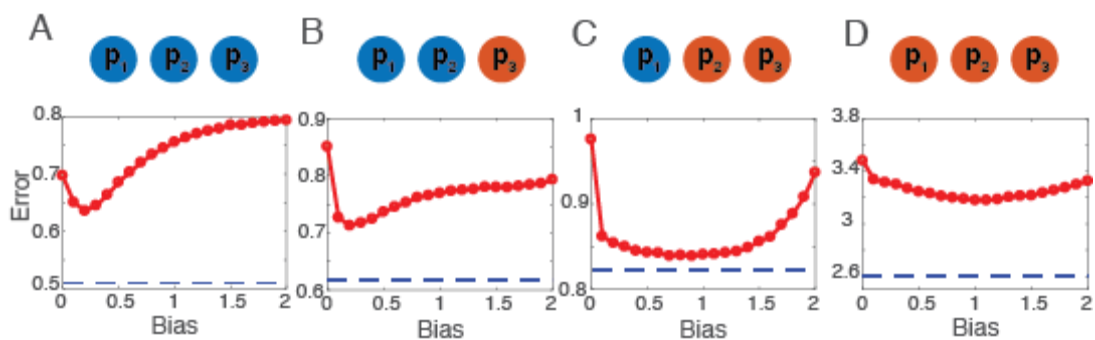

**Figure 3**: **Impact of the social discounting bias**. The average error of the group as a function of the social discounting bias. For different group compositions (i.e. (A) three good performers, (B) two good and one bad performers, (C) one good and two bad performers, and (D) three bad performers) the error curves are non-monotonic. This result indicates that a positive social discounting bias is beneficial to the group performance. Blue dashed lines indicate the average performance of groups that are structured with the optimal weights presented in **figure 2**.

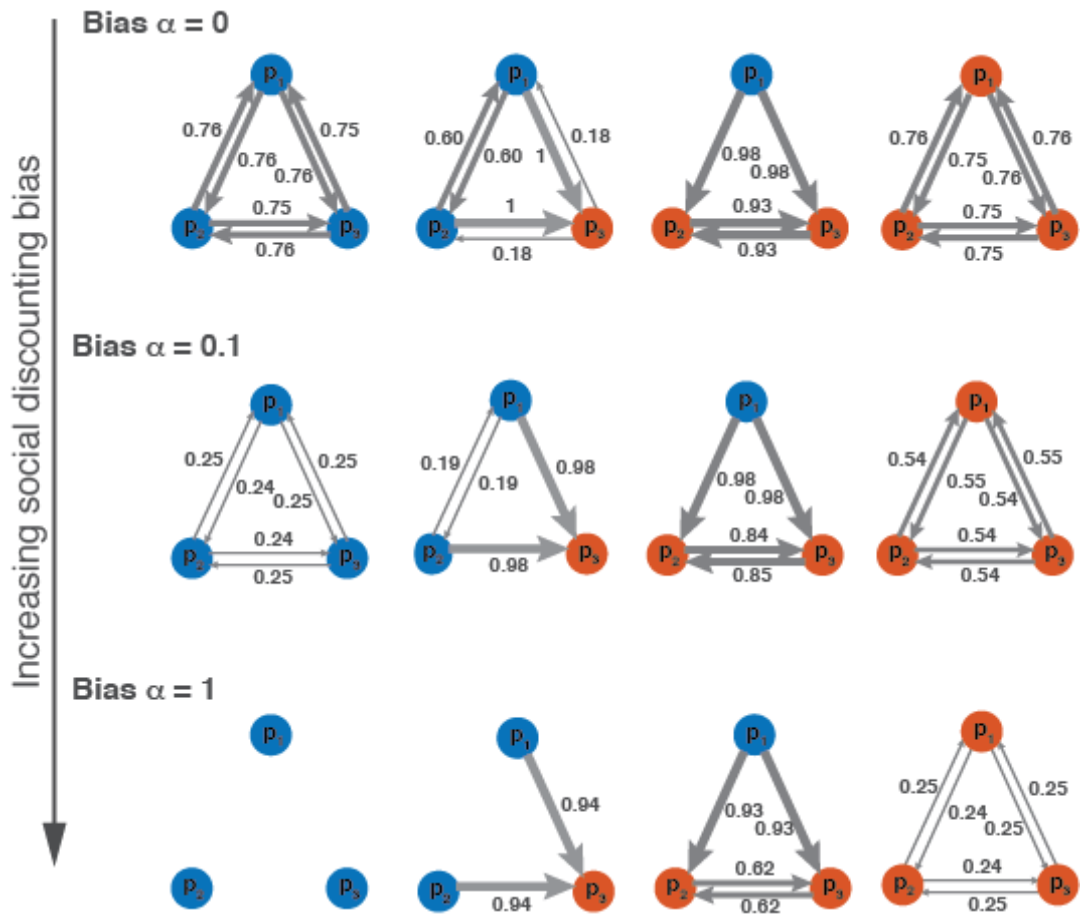

**Figure 4**: **Emerging group structures**. Networks of influence for different group compositions, as they emerged in simulations after t=100 successive discussions. At the end of each discussion, each group member could update the weights given to all others, based on their relative errors in the previous discussion. Over time, the weights gradually converged to the above values. Here, the composition of the group is varied (a good performer is represented in blue whereas a bad performer is represented in red), as well as the social discounting bias $\alpha$. When the majority of group members are good performers (i.e., 3 or 2 blue dots), a small bias of $\alpha = 0.1$ yields the best performances (see figures 3A and 3B) and produces networks of influence similar to the normative ones (see figure 2). When the majority of group members are bad performers (i.e., 2 or 3 red dots), larger bias of $\alpha = 1$ is preferable, producing networks similar to the normative ones.

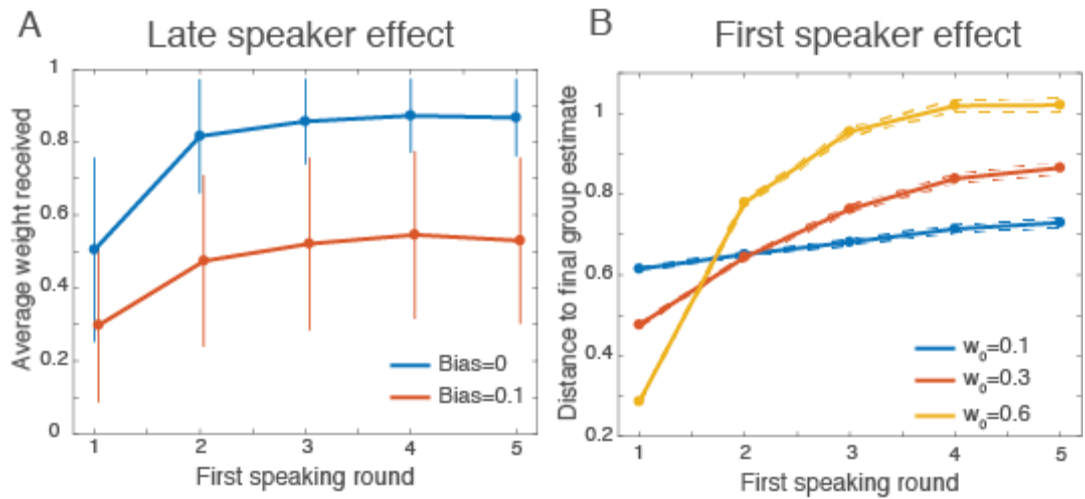

**Figure 5**: **Speaking order effects**. (A) Average weight received by group members after 100 consecutive discussions, as a function of their first speaking round within each discussion. Individuals who speak later during the discussion tend to receive a stronger weight in the long run. The late speaker effect is attenuated by stronger biases. Results are averaged across 1000 simulations. (B) Within a discussion, individuals who speak earlier have a stronger impact on the outcome of that discussion. The first speaker effect is amplified when the weight assigned to the speaker is greater. Results are averaged across 1000 simulations.

| | |
|---|---|
| $x_i^r$ | Estimate of individual *i* at round *r* |
| $x_i^0$ | Initial estimate of individual *i* |
| $N$ | Group size |
| $e_i^r$ | Error associated with the estimate $x_i^r$ |
| $w_{ij}$ | Weight given by individual *i* to the estimates of individual *j*. |
| $\alpha$ | Social discounting bias |
| $\mu, \sigma$ | Mean and standard deviation of the normal distribution from which the individuals' initial estimates are drawn. |
| $N_r$ | Number of speaking rounds in a discussion |
| $N_T$ | Number of discussions. |
| $t$ | Current discussion |
| $w^0$ | Default weight assigned to a stranger |
| $w^*$ | Change of weights between discussions |
| $x_i^*$ | First estimate communicated by i during the discussion. |

**Table 1**: **Model variables and parameters**

**Supplementary figures**

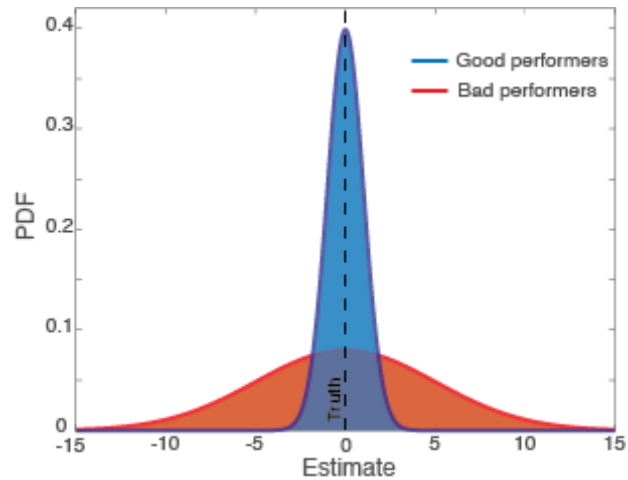

**Figure S6**: **Performance of the agents**. In the simulations, good performers have their initial estimate $x^0$ randomly drawn from a normal distribution with a mean 0 and standard deviation 1 (blue distribution). Bad performers draw their initial estimate $x^0$ from a normal distribution with a mean 0 and standard deviation 5. Estimates are assumed to be normalized such that the truth always equals 0. In such a way, the error **e** associated with a given estimate **x** is simply given by $e = |x|$.

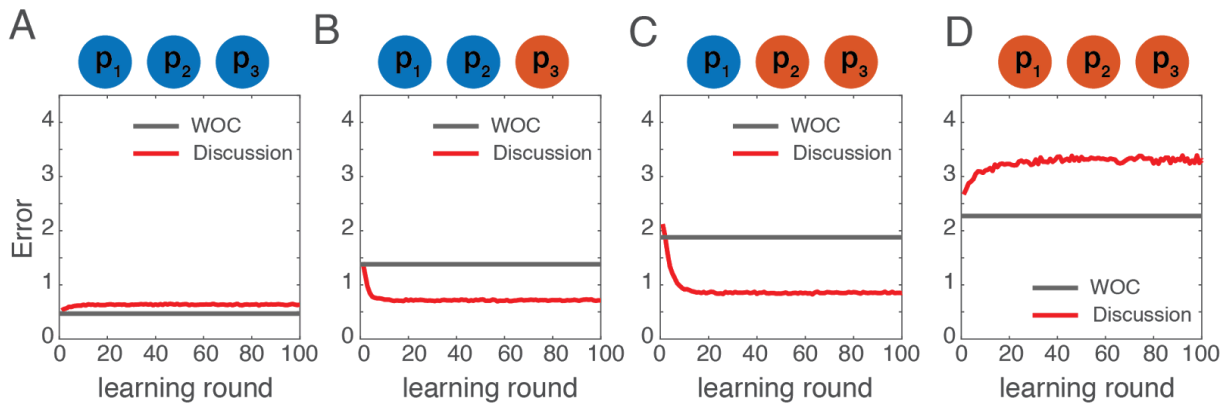

**Figure S7**: **Wisdom-of-the-crowds**. Comparison between the performances of the group discussions and the wisdom-of-the-crowds approach (WOC) for different group compositions and over 100 learning rounds. The WOC is evaluated by measuring the error of the average estimate of the group members before the discussion starts. The WOC does not involve interaction between group members and is therefore identical across all learning rounds for a given group compositions. In contrast, the performance of the group discussions depends on the weights that the group members assigned to one another and therefore change over learning rounds. When all group members are equally skilled (either all good or all bad), the discussion is outperformed by the WOC (in A and D). However, when there exist skill differences within the group (in B and C), the discussion eventually outperforms the WOC

because group members gradually learn to rely on the judgment of their best performers, whereas the WOC weights the judgments of the good and bad performers equally. Results are averaged over 5000 simulations, with $N = 3$ and $\alpha = 0$.